# Nonvolatile Electric Control of Antiferromagnet CrSBr

*Junhyeon Jo[1,*], Samuel Mañas-Valero[2], Eugenio Coronado[2], Fèlix Casanova[1,3],*

*Marco Gobbi[3,4,*], Luis E. Hueso[1,3,*]*

[1] CIC nanoGUNE BRTA, 20018 Donostia-San Sebastian, Basque Country, Spain

[2] Instituto de Ciencia Molecular (ICMol) Universitat de València, Catedrático José Beltrán 2, Paterna 46980, Spain

[3] IKERBASQUE, Basque Foundation for Science, 48013 Bilbao, Basque Country, Spain

[4] Centro de Física de Materiales (CFM-MPC) Centro Mixto CSIC-UPV/EHU, 20018 Donostia-San Sebastián 20018, Basque Country, Spain.

*Corresponding Authors

van der Waals magnets are emerging as a promising material platform for electric field control of magnetism, offering a pathway towards the elimination of external magnetic fields from spintronic devices. A further step is the integration of such magnets with electrical gating components which would enable nonvolatile control of magnetic states. However, this approach remains unexplored for antiferromagnets, despite their growing significance in spintronics. Here, we demonstrate nonvolatile electric field control of magnetoelectric characteristics in van der Waals antiferromagnet CrSBr. We integrate a CrSBr channel in a flash-memory architecture featuring charge trapping graphene multilayers. The electrical gate operation triggers a nonvolatile 200 % change in the antiferromagnetic state of CrSBr resistance by manipulating electron accumulation/depletion. Moreover, the nonvolatile gate modulates the metamagnetic transition field of CrSBr and the magnitude of magnetoresistance. Our findings highlight the potential of manipulating magnetic properties of antiferromagnetic semiconductors in a nonvolatile way.

Manipulating the magnetic properties of materials via electric fields is an attractive alternative to traditional magnetic field control.[1-10] Importantly, it eliminates the need of macroscopic magnetic fields, which are difficult to screen and control at the nanoscale. Moreover, it offers the potential of a reduction in power consumption in spintronic devices by removing the high electrical currents traditionally used for local magnetic field generation or spin torque. Taking a step further beyond the simple electric field control of magnetism, ferroelectric elements can provide both an electric field and the further advantage of nonvolatility as their electric polarization is kept even if a voltage bias is not applied. The combination of ferromagnetic and ferroelectric (or multiferroic) materials has already enabled advanced devices such as multiferroic tunnel junction[11-13] and magnetoelectric spin-orbit.[14-16]

Antiferromagnetic materials have gained prominence in recent years due to several of their unique magnetic properties such as net zero magnetization, absence of stray fields, and ultrafast spin dynamics.[17-19] Some of these properties have been exploited in spintronics, giving rise to phenomena such as antiferromagnetic anisotropic magnetoresistance, anomalous spin Hall effects, and ultrahigh tunneling magnetoresistance.[20-26] However, nonvolatile electric field control in antiferromagnetic materials is still a topic under intense scrutiny.

The reduced dimensionality of atomically thin layered magnetic materials makes them an ideal platform for the manipulation of magnetism either via tailored interfaces or through electric field effects.[27-34] In particular, various studies have demonstrated how the Curie temperature and the magnetic anisotropy of two-dimensional (2D) magnets can be manipulated by electrostatic gating. Furthermore, the discovery of ferroelectric order in 2D materials[35-39] has ignited the concept of utilizing them for the nonvolatile control of the magnetic properties in 2D ferromagnets.[40-42] Recent findings have shown the modification of magnetic characteristics in layered antiferromagnets through electrostatic fields,[29, 30, 34] although the demonstration of nonvolatile gate control has not been achieved so far.

Here, we show that a nonvolatile electric field controls both the electronic and magnetic properties of a van der Waals semiconducting antiferromagnet CrSBr. We make use of a flash-memory architecture, consisting of a charge trapping graphene multilayer (Gr) separated from the CrSBr active channel through a tunneling barrier of hexagonal boron nitride (hBN), with a ferroelectric $CuInP_2S_6$ (CIPS) layer as a top gate dielectric. In this heterostructure, the charging/discharging of the Gr layer induces a nonvolatile 20-fold modification in the electrical resistance of CrSBr which further varies depending on the magnetic state of CrSBr. Moreover, the nonvolatile electrical gate modifies the magnetic anisotropy, as evidenced by nonvolatile changes in both magnetoresistance (MR) and magnetic critical field. Our findings highlight the potential of manipulating magnetic characteristics in 2D antiferromagnetic semiconductors by means of nonvolatile electric fields.

CrSBr is a layered antiferromagnet characterized, at zero magnetic field, by ferromagnetic ordering within each layer and antiferromagnetic coupling between neighboring layers.[43-45] Notably, the interlayer antiferromagnetic coupling can be overcome by the application of a magnetic field which switches the alignment of the magnetization from an antiparallel (AP) to a parallel (P) state. Figure 1a shows MR curves measured in a pristine CrSBr (8 layers) flake on a $Si/SiO_2$ substrate in response to an applied magnetic field. In this study, we focus on the magnetic field applied along the *b*-axis which is the magnetic easy axis of CrSBr. When the magnetic field is lower than a critical field of around 4 kOe, the magnetization of adjacent layers is in the AP state. As the field increases above the critical value, the AP alignment is switched to the P state. Correspondingly, a sharp change in the resistance is observed, as the AP state is characterized by a higher resistance than the P state.[43] The critical fields present a small shift depending on the field sweep directions, which may originate from spin torque or bias effect exerted during the spin-flip process of the layers.[44]

Figure 1b displays the electrical gate ($V_G$)-dependent transport properties of CrSBr measured using a highly doped-Si substrate as a bottom gate and a 300 nm $SiO_2$ film as a dielectric layer. A significant drop (increase) in resistance is observed at positive (negative) $V_G$ in agreement with the *n*-type semiconducting nature of CrSBr.[46] Here, we note that a slight hysteretic loop is detected during different $V_G$ sweep directions, which comes from the charging effect of the $SiO_2$ dielectric layer.[47]

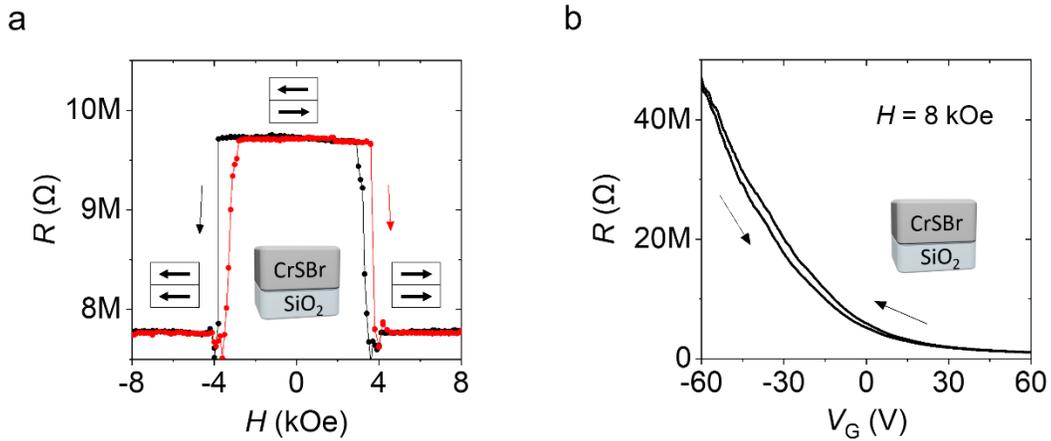

**Figure 1.** Magnetic field and electrical gate response on antiferromagnetic semiconductor CrSBr. (a) Magnetotransport in a CrSBr layer. A magnetic phase transition between an antiparallel and parallel alignment of the magnetization of adjacent layers occurs through a spin-flip process. (b) Resistance as a function of an electrical gate ($V_G$), applied through a $SiO_2$ bottom dielectric layer. A significant decrease in resistance is observed for positive $V_G$, indicating *n*-type semiconducting behavior. The measurements were performed at 10 K.

Combining the magnetic properties of CrSBr with its semiconducting nature offers the possibility to achieve nonvolatile electrical gate control over both the electrical resistance and the magnetic anisotropy of CrSBr. Figure 2a shows a schematic image of the full heterostructure composed of CrSBr and Gr layers as well as intermediate hBN and top CIPS (*i.e.* CrSBr/hBN/Gr/CIPS). Here, the Gr multilayer, isolated from the CrSBr channel by the hBN, is used as a charge trapping layer in the floating gate configuration.[48] The role of this layer is to charge/discharge carriers which gives nonvolatility of the electric and magnetic tuning into the CrSBr. When a positive $V_G$ is applied during the negative direction $V_G$ sweep, charge carriers (electrons) in the CrSBr channel deplete as the electrons tunnel into the Gr trapping layer through Fowler-Nordheim tunneling in the hBN layer, enhancing the resistivity of the CrSBr channel (Figure 2c). Conversely, applying a negative $V_G$ in the opposite direction sweep results in electron accumulation in the CrSBr, as the stored electrons in the Gr layer tunnel back to the CrSBr, leading to the lower resistivity of the CrSBr (Figure 2d). For this reason, an hBN layer of between 7 nm and 8 nm is used in our study to avoid direct tunneling which degrades the charge trapping process. On the other hand, the top CIPS layer plays a role as a gate dielectric to modulate overall band alignment for optimal Fowler-Nordheim tunneling process in this study. In addition, the manipulation of charge carriers by electric fields can change the hybridization between itinerant charge carriers and magnetic moments[1, 49] through *s-d* exchange coupling with magnetic ions or defects in CrSBr lattices.[44] Accordingly, a nonvolatile electrical gate not only affects CrSBr electrical conductivity, but also its magnetic anisotropy and resulting magnetotransport characteristics.

The heterostructure was fabricated using a polymer-assisted dry transfer method in an Ar-filled glovebox, and it was placed on top of prepatterned Au electrodes fabricated on a Si/SiO$_2$ substrate (Figure 2b). A voltage is applied along the *a*-axis of CrSBr between a drain and source electrode, while a magnetic field is applied along its *b*-axis. As shown in Figure 1a, this configuration yields sharp magnetic switching due to the magnetic anisotropy of CrSBr.[43, 50, 51] An additional Gr layer is placed on top of the CrSBr/hBN/Gr/CIPS heterostructure as a top gate electrode.

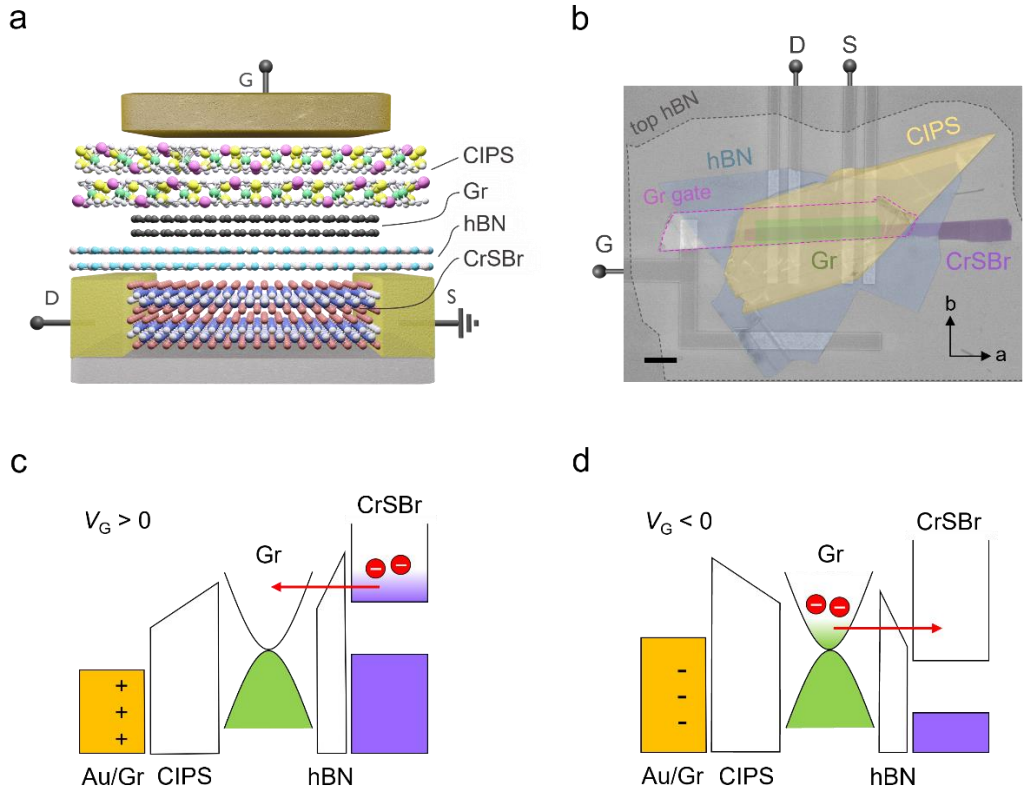

**Figure 2.** Working principle of nonvolatile gate control on the antiferromagnetic CrSBr in a CrSBr/hBN/Gr/CIPS heterostructure. (a) Schematic image of the heterostructure. The layers of hBN/Gr are located between the CrSBr and CIPS, with the Gr layer isolated from the CrSBr channel. (b) Scanning electron microscopy image of the fabricated heterostructure. A voltage is applied along the *a*-axis, and the longitudinal current is collected using a drain (D) and source (S) electrode. A magnetic field is applied along the *b*-axis. An additional Gr layer (pink dotted line) on top of the heterostructure is used as a top gate electrode (G), and all the structure is encapsulated by a top hBN layer (grey dotted line). (c,d) Schematic band diagrams of the Gr-based floating gate system. Application of a positive gate ($V_G$) during the negative direction sweep leads to tunneling electrons into the Gr trapping layer through the Fowler-Nordheim tunneling process in the hBN layer. The opposite gate process results in electron tunneling into the CrSBr from the Gr trapping layer.

Figure 3a shows the temperature dependence of resistance, $R(T)$, of an antiferromagnetic CrSBr (8 layers) recorded while ramping up the temperature after cooling the device with different applied top electrical gate voltage ($V_G$ = +4 V, 0 V, and −4 V). The $R(T)$ curves measured for the three electric field cooling (EFC) conditions are markedly different, even if the $V_G$ is set to 0 V in the three cases during the resistance measurement. In particular, the $R(T)$ with EFC at $V_G$ = 0 V (black curve) shows the expected semiconducting behavior for CrSBr (see further transport analysis of this semiconductor in Figure S1). For the case of EFC at $V_G$ = +4 V, a lower resistance is observed for all temperatures as electrons are accumulated in the CrSBr channel, indicating a remanent *n*-type doping in the CrSBr (red curve). In contrast, the resistance is significantly higher after EFC at $V_G$ = −4 V (blue curve) as electrons are depleted. The dependence of the resistance with the EFC procedure shows a clear nonvolatile electrical gating effect.

The change in resistance as a function of an applied electrical gate, measured at 10 K, displays a large nonvolatile hysteresis (Figure 3b). At $V_G$ = +4 V, a low resistance state is recorded. As $V_G$ is swept towards the negative direction, the resistance rapidly increases and saturates at $V_G$ = +1.8

V in a high resistance state (electron depletion in the CrSBr channel), which is maintained down to $V_G = -4$ V. Conversely, when $V_G$ is swept from negative to positive voltages (from $V_G = -4$ V to +4 V), the resistance decreases and quickly reaches a saturated state at $V_G = -2$ V (electron accumulation). The observed inverted-hysteretic gate window indicates charging and discharging of carriers through the Gr layer (see $V_G$ range dependence in Figure S2). The stored electrons in the Gr multilayer is estimated to $7.1 \times 10^{12}$ cm$^{-2}$ analogous to reported Gr-based floating gate systems.[48, 52].

Figure 3c,d displays the retentive and dynamic gate control behavior in our device. To confirm its nonvolatile feature, retention measurements were performed over time in the accumulation and depletion state of the CrSBr, separately. For the accumulation state in the CrSBr, $V_G = -5$ V was applied for 2 s and $V_G$ was turned off, leading to the low resistance state. Subsequently, the resistance was measured at $V_G = 0$ V for an hour (accumulation, the red curve in Figure 3c). A slight increase of resistance is detected at the beginning of the measurement, but it saturates soon after. For the depletion state, $V_G = +5$ V was imposed following the same procedure (depletion, the blue curve in Figure 3c). The detected resistance values are stable over time. Figure 3d displays the dynamic behavior for the electron accumulation/depletion state. The low resistance is measured while applying $V_G = +5$ V in which electrons are accumulated in the CrSBr channel. Next, applying $V_G = 0$ V drives the device to the saturated high resistance state, as the electrons are trapped in the Gr layer by the Fowler-Nordheim tunneling barrier of hBN. A following application of $V_G = -5$ V retains the high resistance state, but the continuous application of $V_G = 0$ V allows the captured electrons to move back to the CrSBr layer from the Gr trapping layer. This carrier transfer triggers

a gate-hysteresis window with the high and low resistance state in the antiferromagnetic state of CrSBr. We note that 20-fold difference is measured between the two resistance states of CrSBr.

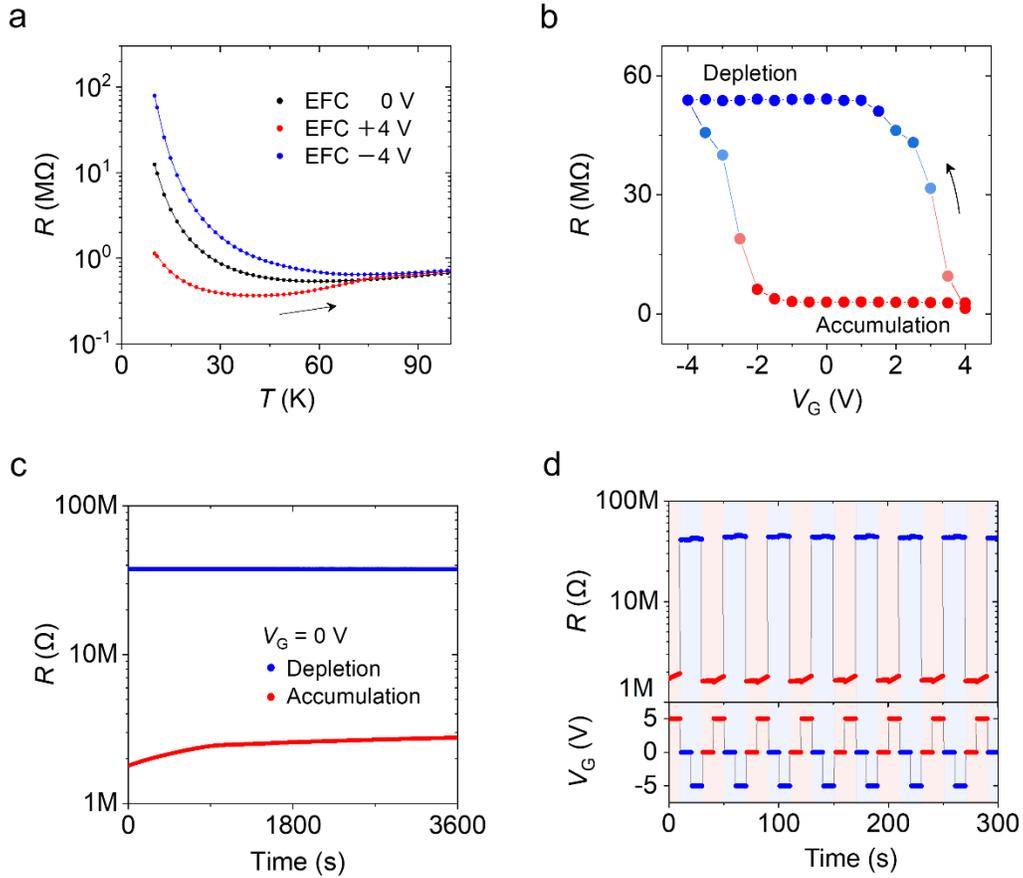

**Figure 3.** Nonvolatile gate control on the antiferromagnetic CrSBr in a CrSBr/hBN/Gr/CIPS heterostructure. (a) Temperature dependence of the resistance measured with zero gate voltage ($V_G$ = 0 V) after an electric field cooling (EFC) procedure with $V_G$ = 0 V (black), $V_G$ = +4 V (red), and $V_G$ = −4 V (blue). (b) Gate-induced inverted-hysteresis loop for the antiferromagnetic state of CrSBr at 10 K. Different resistances are measured in the CrSBr channel, corresponding to the electron accumulation and depletion states in the CrSBr. (c) Retention characteristics for the electrons accumulation and depletion state. The resistance is measured at $V_G$ = 0 V after applying

a positive ($V_G$ = +5 V, blue) or negative ($V_G$ = −5 V, red) gate pulse with a duration of 2 s. (d) Dynamic characteristics recorded by applying gate pulse cycles. Identical resistance values are detected for $V_G$ = +5 V and −0 V and for $V_G$ = +0 V and −5 V, corresponding to the gate-hysteresis loop in (b).

Next, we show that the electric field control also modifies the magnetic properties of CrSBr in a nonvolatile way. Figure 4a displays nonvolatile hysteresis loops in resistance which depend on the magnetic state of CrSBr; the AP state under $H$ = 0 kOe and the P state under $H$ = 8 kOe. For both the AP and P state, the low resistance is recorded when electrons are accumulated in CrSBr. Conversely, the high resistance state is shown in the electron depletion region. While the overall hysteretic behavior is similar for the AP and P state, the resistances of the two saturated states are markedly different. In this way, our nonvolatile CrSBr device shows four stable resistance states, addressable by a combination of a nonvolatile electric field effect and a magnetic field. As temperature rises and approaches the critical temperature of CrSBr, which is approximately 140 K, the difference in the hysteresis loops of the two magnetic states diminishes (Figure S3). Even though the hysteresis loops can be observed at 200 K, much above the critical temperature, there is no discernible distinction between the two magnetic states at such temperature (Figure S4).[53, 54]

Figure 4b displays MR curves measured in the two remanent resistance states, where MR is defined as $MR\ (\%) = (R(H) - R(H=0)) / R(H=0) \times 100$. To facilitate this comparison, we only present results obtained during a magnetic field sweep in the negative direction. We define $V_G$ = +0 V and $V_G$ = −0 V as an electrical gate state of $V_G$ = 0 V which is reached after a gate sweep from a positive and a negative value, corresponding to ② and ④ in Figure 4a, respectively. At $V_G$ = +4 V and $V_G$ = −0 V which correspond to the accumulation state of the CrSBr (① and ④ in Figure 4a), MR reaches around 27 %. However, for $V_G$ = −4 V and $V_G$ = +0 V, corresponding to the depletion state (③ and ② in Figure 4a), the magnitude of MR reduces to 16 % (see a full gate-dependent MR curve in Figure S5). Thus, the MR of CrSBr is also modified by the remanent state of carrier charging/discharging. Another noteworthy aspect associated with remanent electric fields is a change of the critical field required for the AP-to-P state switching in CrSBr. In particular, the MR traces display lower switching fields for the accumulation state (① and ④) than for the depletion state (③ and ②). The critical fields change from −3900 Oe to −3650 Oe in a negative field region and from +3750 Oe to +3300 Oe in a positive magnetic field region (dotted lines in the magnified graph on the right panel in Figure 4b). Since this variation is linked to the remanent state of the charging/discharging, the switching fields also display a clear hysteretic behavior (Figure S6). These findings demonstrate that the nonvolatile electric field affects not only the electrical but also the magnetic properties of CrSBr. We ascribe this effect to carrier-mediated exchange coupling or variations in electronic band dispersion associated with charge carriers and magnetic ions in CrSBr.

Figure 4c,d displays the overall trend of electrical gate-dependent MR (with a gate step of 0.5 V) as color maps for both directions of a gate sweeping. For every $V_G$, the P state exhibits a lower resistance than the AP state. When $V_G$ is swept from positive to negative values (Figure 4c), a sharp transition takes place near $V_G$ = +3 V from the low resistance state to the high resistance state. Conversely, when $V_G$ is swept in the opposite direction, a reduction in resistance is observed

near $V_G = -2.5$ V (Figure 4d). These features correspond to the gate-hysteresis loops in Figure 4a related with the magnetic states of CrSBr. Thus, the MR analysis offers compelling evidence of the nonvolatile gate control over the magnetic anisotropy, substantiated by the metamagnetic transition of CrSBr.

A control sample with a simpler hBN gate dielectric layer (CrSBr/hBN/Gr/hBN) was fabricated for comparison to the complete CIPS-based structure (CrSBr/hBN/Gr/CIPS). The hBN-gate structure presents similar characteristics with the CIPS structure in gate- and magnetic field-dependent measurement. However, it shows poorer performance such as a narrower gate-induced inverted-hysteretic window and a reduced difference in the MR (Figure S7). Therefore, our data shows that the use of CIPS instead of hBN as a top gate dielectric improves the efficiency of the floating gate. We ascribe the better performance of the CIPS device to a more suitable band alignment for the CrSBr/hBN/Gr junction, which leads to an optimum Fowler-Nordheim tunneling, while applying an electrical gate bias through the CIPS layer that could be further enhanced in its domains due to the interfaced Gr layer.[55]

Another noteworthy approach to realize nonvolatile electric control on antiferromagnetism could be ferroelectric switching by interfacing a ferroelectric CIPS layer and an antiferromagnet CrSBr. Indeed, a CrSBr/CIPS heterostructure shows a clear ferroelectric gate response in both the AP and P magnetic state of the CrSBr, showing 12 % resistance difference between in the positive and negative ferroelectric polarization state at 100 K (Figure S8a). Moreover, the change of MR between the AP and P state is recorded as 3.6 % and 4.6 % in the positive and negative ferroelectric polarization state, respectively (Figure S8d-f). However, the ferroelectric response from the CIPS disappears when temperature decreases (Figure S8b,c). We ascribe this disappearance to not fully polarized ferroelectric domains of the CIPS due to limited displacement of Cu ions at low temperatures.[56]

We have demonstrated nonvolatile gate control of magnetism and magnetotransport in the antiferromagnetic semiconductor CrSBr. We used a floating gate structure based on a Gr charge trapping layer with the assistance of a ferroelectric CIPS layer; the resulting device displays a sharp electrical gate-induced inverted-hysteresis characterized by a remarkable 200 % resistance difference in the antiferromagnetic state of CrSBr. Moreover, the nonvolatile electric field introduces a change in the MR magnitude and in the metamagnetic switching field of the CrSBr approximately 10 %. This indicates that not only the CrSBr magnetoelectrical resistance is retentively modified, but also its magnetic anisotropy can be controlled in a nonvolatile way. We envision that this work lays the foundation for the integration of nonvolatile gating architectures with antiferromagnetic materials, representing a significant milestone in the evolution of next-generation spintronics devices.

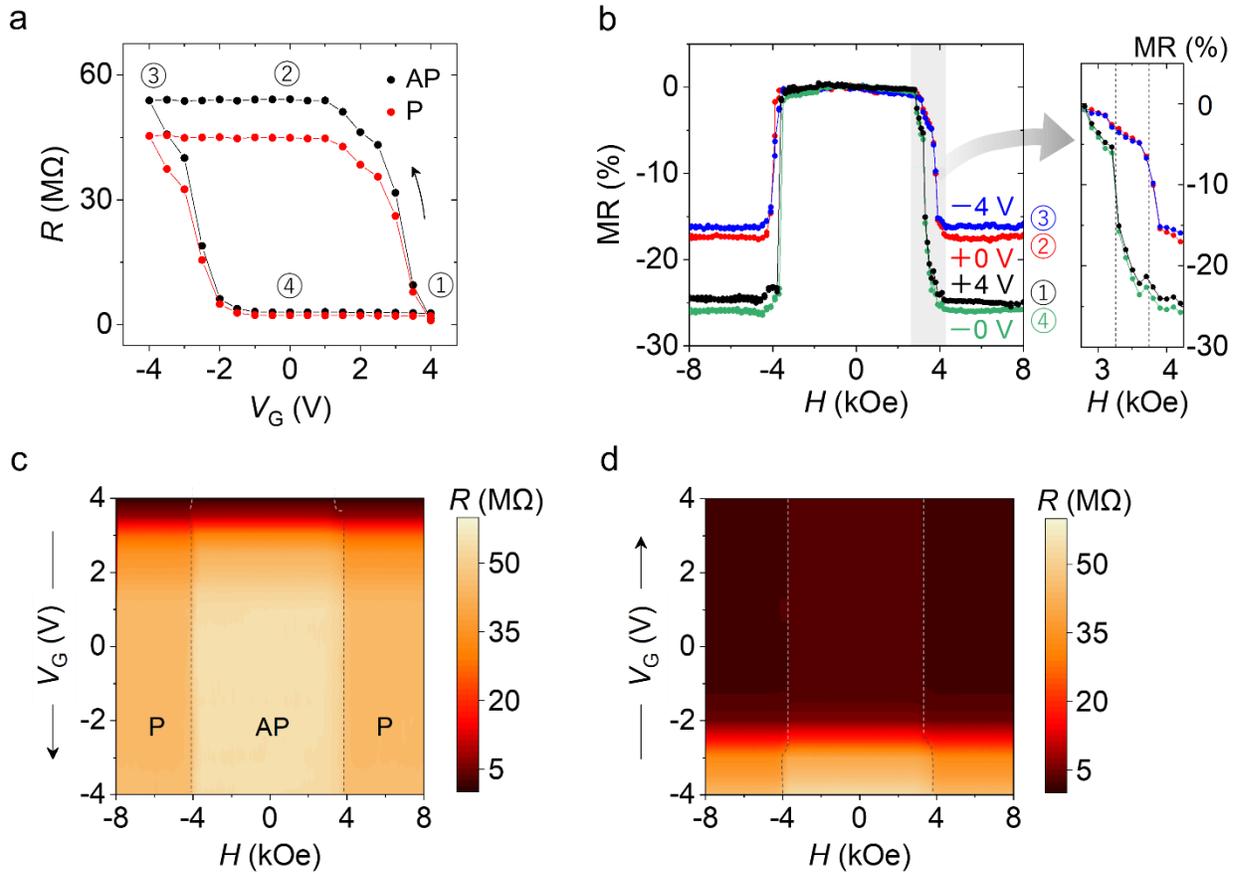

**Figure 4.** Electrical gate control on magnetotransport of CrSBr in a CrSBr/hBN/Gr/CIPS heterostructure. (a) Gate-hysteresis loop for the different magnetic states of CrSBr at 10 K. In the antiparallel (AP) and parallel (P) magnetic state, distinct hysteresis loops appear, while two different states in resistance are detected corresponding to electron accumulation (① and ④) and depletion (② and ③) in the CrSBr. (b) Magnetoresistance (MR) under the accumulation and depletion state. The magnitude of the MR and the critical field of the metamagnetic transition in CrSBr change upon the remanent charging states, indicating nonvolatile gate control over magnetic anisotropy. Numbers indicate the corresponding states in (a). A magnified image for the positive critical field region is shown in the right panel and the dotted lines indicate the critical field for each state. (c,d) MR color maps depending on the direction of a gate sweep. The magnetic

transition of CrSBr appears near 4 kOe which is identical value to the one observed in (b). A negative gate sweep (c) shows a significant resistance change near $V_G$ = +3 V, while a positive sweep (d) shows the change around $V_G$ = −2.5 V.

**Methods**

Device Fabrication: Designed electrodes were prepared on a Si/SiO$_2$ substrate using electron beam lithography. The electrodes consisted of 2 nm of Ti and 10 nm of Au. The width of the electrodes was 1 μm, and the channel gap between them was 4 μm. Bulk 2D crystals of CIPS and hBN were purchased from HQ graphene. CrSBr crystals were synthesized by chemical vapor transport using direct reaction of components.[53] 2D heterostructures were fabricated in an Ar-filled glovebox using mechanical exfoliation and a polymer-assisted dry transfer method. 8 layers of CrSBr was used in this study, and a hBN layer between 7 nm and 8 nm was used as a Fowler-Nordheim tunneling barrier for a CIPS gate-controlled structure and a hBN gate-controlled structure, respectively. Additional Gr layer was used for the top gate electrode and the heterostructure was encapsulated by a hBN layer. The thickness of 2D flakes was first cross-checked by optical contrast and atomic force microscopy measurement, and the optical contrast was used during the fabrication in the glovebox.

Electrical measurement: A fabricated device was mounted in the sample stage of the PPMS (Physical Property Measurement System, Quantum Design) with a horizontal rotator. Electrical measurement was performed with Keithley 2636. A source voltage of 0.5 V was applied to a channel CrSBr, and a gate voltage through top gate electric layers was imposed. A leakage current from a gate bias was less than 0.1 nA. A constant rate for a magnetic field sweep, 100 Oe/s, and for a temperature sweep, 2 K/min, were used for data collection. The EFC procedure was conducted as applying a constant source voltage of 0.5 V and a target gate voltage during cooling from 200 K to 10 K.

**Conflict of Interest**

The authors declare no competing financial interest.

**Acknowledgements**

This work was supported by the Spanish MCIN/AEI under Projects PID2020-117152RB-100, PID2021-128004NB-C21, TED2021-130292B-C42, and PID2021-122511OB-I00. This work was also supported by the FLAG-ERA grant MULTISPIN, by the Spanish MCIN/ AEI with grant number PCI2021-122038-2A. This work was supported by CEX2019-000919-M and CEX2020-001038-M /AEI /10.13039/ 501100011033 under the María de Maeztu Units of Excellence Program. Funding from the European Union's Horizon 2020 Research and Innovation Program under Project SINFONIA, Grant 964396 as well as ERC AdG Mol-2D 788222 and ERC-2021-StG-101042680 2D-SMARTiES are acknowledged. The work was supported by Generalitat Valenciana (PROMETEO Program, PO FEDER Program IDIFEDER/2018/061 and CDEIGENT/2019/022). This study forms part of the Advanced Materials program and was supported by MCIN withfunding from European Union NextGenerationEU (PRTR-C17.I1), from the Basque country and from Generalitat Valenciana. M.G. acknowledges support from the "Ramón y Cajal" Program by the Spanish MCIN/AEI (grant no. RYC2021- 031705-I). J.J. acknowledges the funding from the Ayuda FJC2020-042842-I financiada por MCIN/AEI/10.13039/501100011033 y por la Unión Europea NextGenerationEU/PRTR.

Supporting Information

# Nonvolatile Electric Control of Antiferromagnet CrSBr

*Junhyeon Jo[1,*], Samuel Mañas-Valero[2], Eugenio Coronado[2], Fèlix Casanova[1,3], Marco Gobbi[4,*], Luis E. Hueso[1,3,*]*

[1] CIC nanoGUNE BRTA, 20018 Donostia-San Sebastian, Basque Country, Spain

[2] Instituto de Ciencia Molecular (ICMol) Universitat de València, Catedrático José Beltrán 2, Paterna 46980, Spain

[3] IKERBASQUE, Basque Foundation for Science, 48013 Bilbao, Basque Country, Spain

[4] Centro de Física de Materiales (CFM-MPC) Centro Mixto CSIC-UPV/EHU, 20018 Donostia-San Sebastián, Basque Country, Spain.

*Corresponding Authors

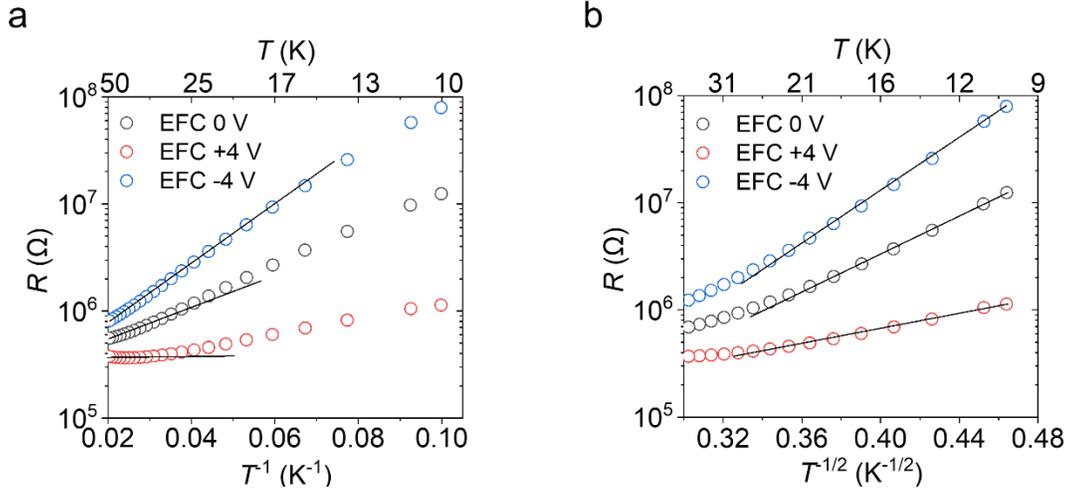

**Figure S1**. Transport behavior of antiferromagnetic semiconductor CrSBr at low temperatures. Data analysis for (a) the Arrhenius plot and (b) the variable range hopping (VRH) for 2D. In the temperature region between 50 K and 30 K, the curves fit to the Arrhenius plot, but the 2D-VRH governs the temperature range between 25 K and 10 K. The transition between the Arrhenius region and VRH region varies on the applied gate during cooling as around 30 K for the EFC +4 V case and 20 K for the EFC −4 V case, which could be inferred that the VRH dominates the transport at low temperature as more electrons exist in the CrSBr under the positive gate which move to nearby hopping sites.

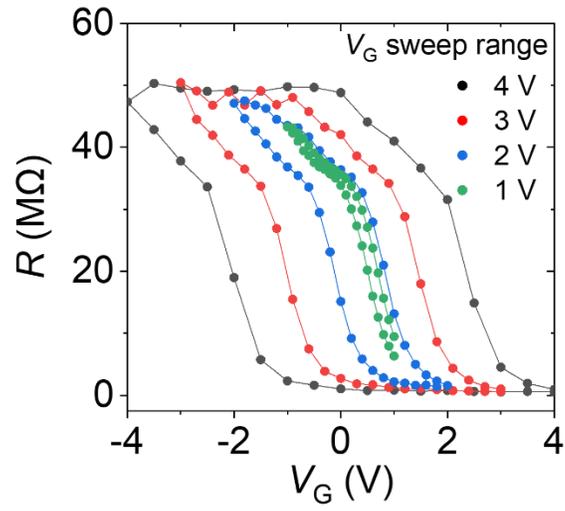

**Figure S2**. Gate-hysteresis loops of the antiferromagnetic state of CrSBr in a CrSBr/hBN/Gr/CIPS heterostructure at 10 K. The hysteresis loop starts to be saturated after a gate of 2.5 V.

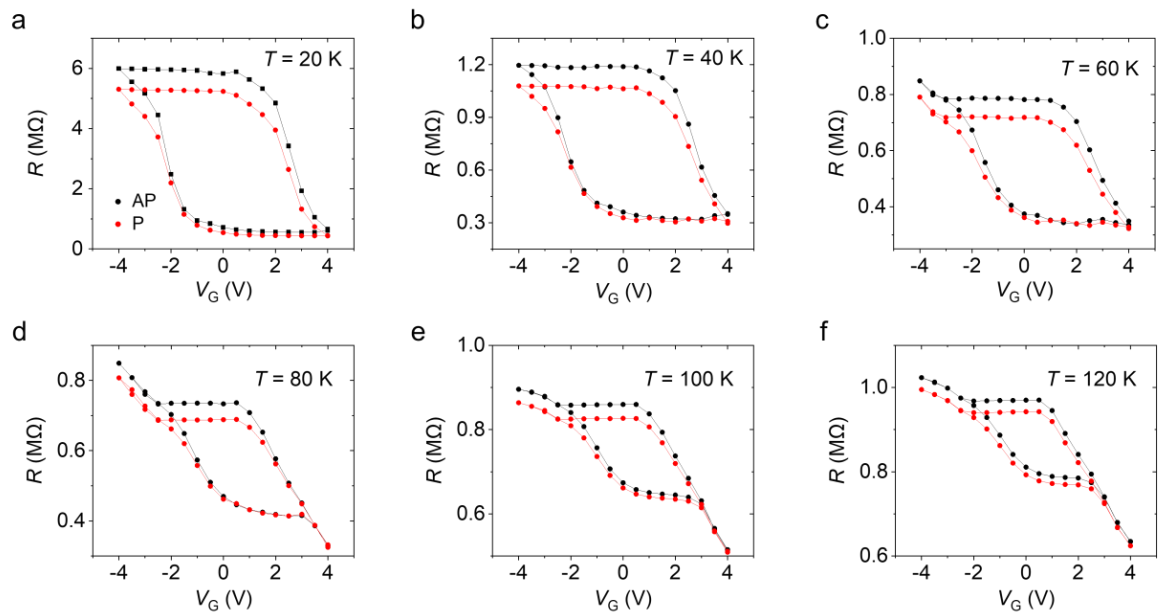

**Figure S3**. Gate-hysteresis loop for the two magnetic states of CrSBr at different temperatures in the CrSBr/hBN/Gr/CIPS device. Resistances curves for the antiparallel (AP) and parallel (P) state are clearly distinguished in the gate-hysteresis loops, but as temperature rises and reaches to the Neel temperature of CrSBr (around 140 K), the difference gets smaller.

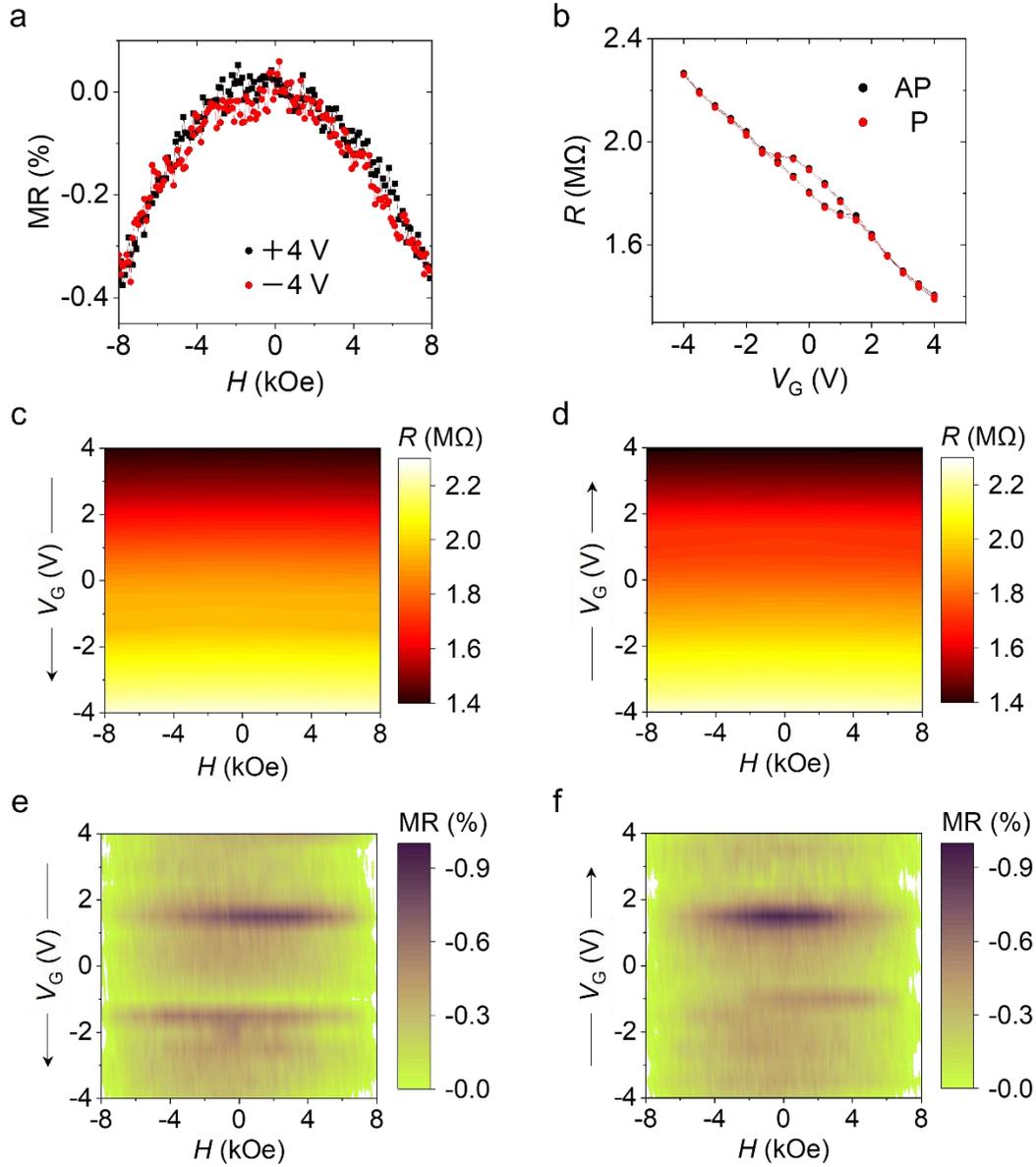

**Figure S4**. Magnetotransport analysis in the CrSBr/hBN/Gr/CIPS heterostructure at 200 K. (a) MR curves with a gate of +4 V and −4 V. There is no discernable difference for different gate polarity. (b) Gate-hysteresis loop measured at a field of 0 kOe and 8 kOe. Even though there is a clear hysteresis, no difference is observed for the two magnetic states of CrSBr; the antiparallel state (AP) and parallel state (P). (c,d) MR and (e,f) MR ratio color maps for different directions of a gate sweep. Identical features in the MR curves are observed for both sweep directions.

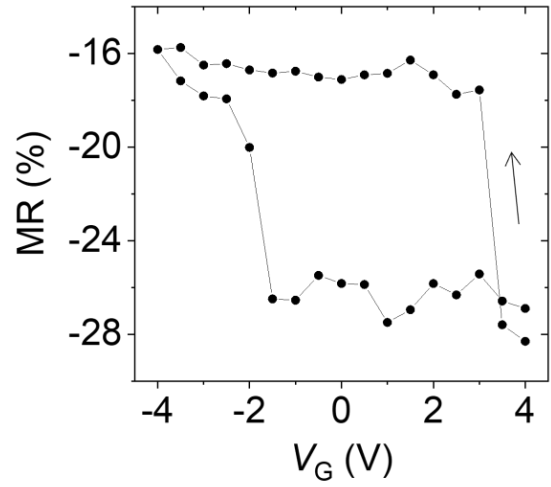

**Figure S5**. MR ratio recorded at different applied gates. Two distinct MR states are recorded, in agreement with the graphs shown in Figure 4b. The black line is a guide to the eye.

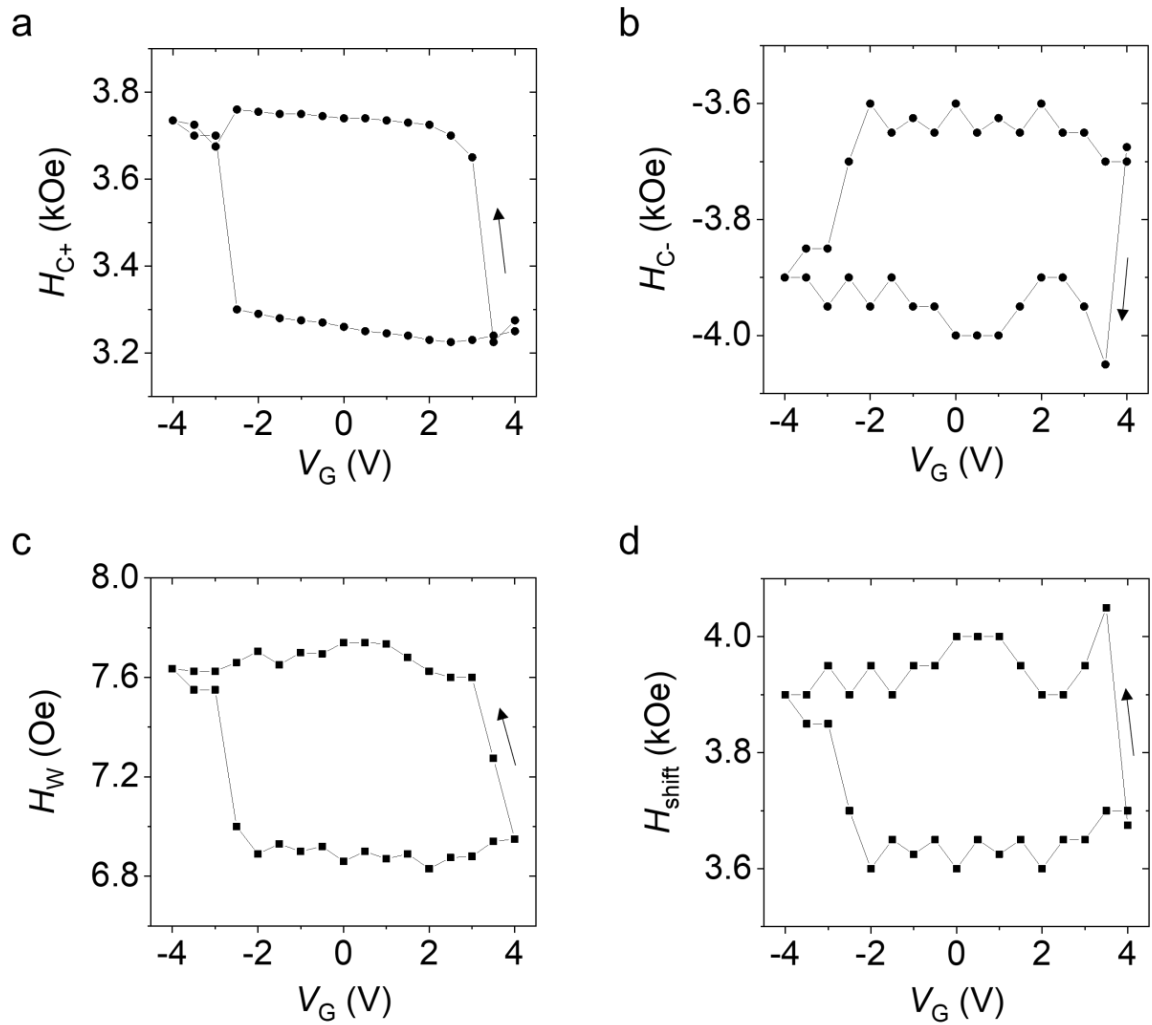

**Figure S6**. Parameters of MR curves at different applied gates. Hysteretic behaviors are observed in the positive critical field (a), negative critical field (b), width (c), and shift in the MR loops (d) from Figure 4b-d. The black lines are guides to the eye.

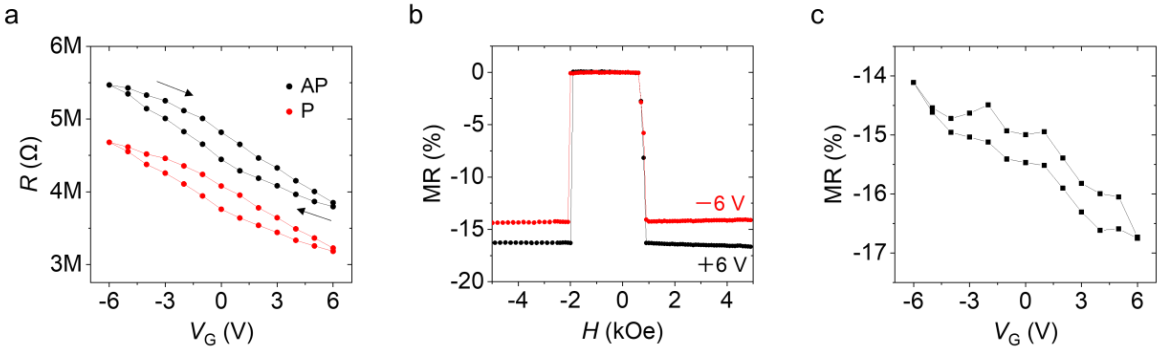

**Figure S7**. Magnetotransport of a CrSBr/hBN/Gr/hBN heterostructure at 10 K. (a) Inverted gate hysteresis for the antiparallel (AP) and parallel (P) state of CrSBr. A gate hysteresis with *n*-type transport is observed, but without two distinct remanent resistance states. (b) MR measured at different applied electric fields. Magnetoresistance (MR) recorded with a positive gate displays a larger ratio as compared to the negative gate case. (c) Gate dependence of the MR ratio. A hysteretic behavior is observed, but without two well-saturated states, similarly to (a).

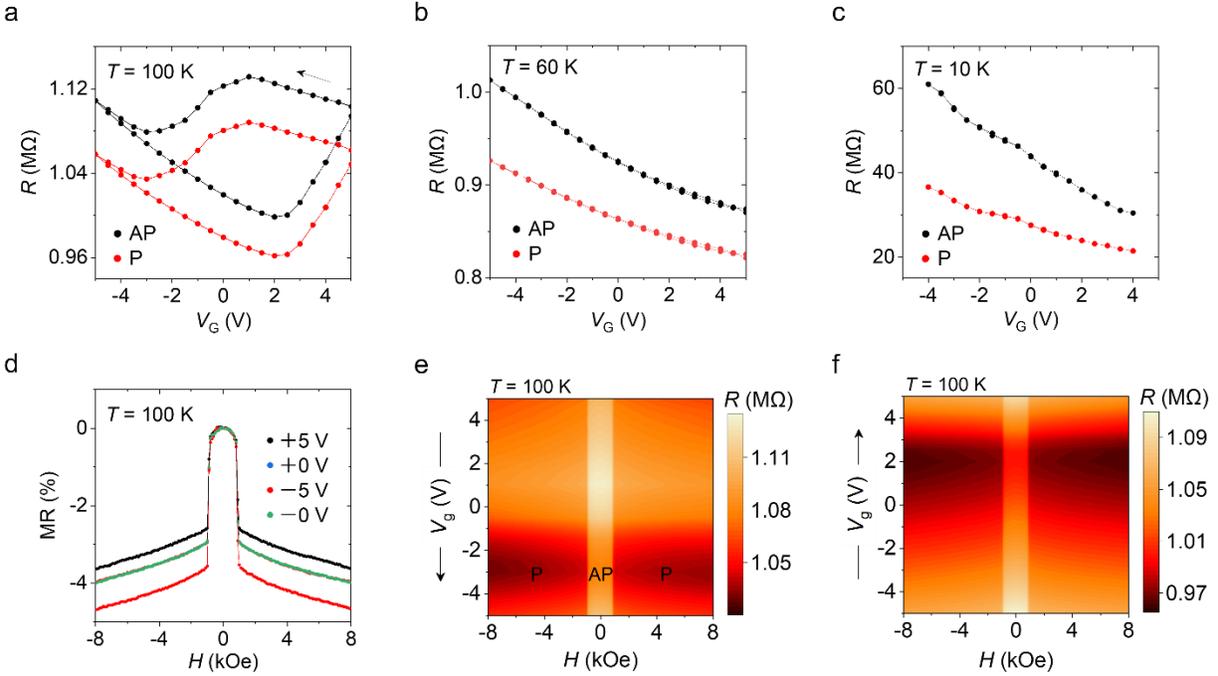

**Figure S8**. Magnetotransport analysis in a CrSBr/CIPS heterostructure. (a-c) Gate-hysteresis loop measured in the antiparallel (AP) and parallel (P) magnetic state of CrSBr at 100 K, 60 K, and 10 K, respectively. Clear ferroelectric windows are observed at 100 K, but it still contains linear background resistances stemming from partial paraelectric domains of the CIPS layer. The ferroelectric features disappear at the temperature below 60 K due to limited displacement of Cu ions. (d) MR curves with a gate of +5 V, +0 V, −5 V, and −0 V at 100 K. Even though two different MR features are detected at +5 V and −5 V, they do not appear the remanent feature at a gate of +0 V and −0 V due to not fully polarized CIPS domains. (e,f) MR ratio color maps for different directions of a gate sweep. Summarized electrical gate and magnetic field responses are shown that are identical to (a) and (d).